\documentclass[twoside]{dis09}
\usepackage[latin1]{inputenc}
\usepackage[dvips]{graphicx,epsfig,color}
\usepackage{wrapfig,rotating}
\usepackage{amssymb,amsmath,array}

\begin{document}
\title{Prompt-photon production in DIS}

\author{Matthew Forrest \\ On behalf of the ZEUS collaboration\cite{url}.
%
%
\vspace{.3cm}\\
%
University of Glasgow - Department of Physics and Astronomy \\
Faculty of Physical Science, University of Glasgow, G12 8QQ - UK
}

\maketitle

\begin{abstract}
Prompt-photon cross sections in deep inelastic $ep$ scattering were
measured with the ZEUS detector at HERA using an integrated luminosity
of $320\mathrm{pb}^{-1}$. Measurements of differential cross sections are presented for inclusive prompt-photon production as a
function of $Q^2$, $x$, $E_T^\gamma$ and $\eta^\gamma$. Perturbative QCD
predictions and Monte Carlo predictions are compared to the measurements.
\end{abstract}

\section{Introduction}
\label{sec-int}

In the study of high energy collisions involving hadrons, events in
which isolated high-energy photons (often referred to as \emph{prompt photons} if originating from quarks) are observed provide a direct probe
of the underlying parton process, since the emission of these 
photons is  unaffected by parton hadronisation. Isolated high-energy
 photon production has been studied in hadronic
collisions\cite{hadexp}
and $ep$ collisions in both the
photoproduction~\cite{prph_php}
and deep inelastic scattering (DIS)~\cite{prph_dis} regimes.
The present results are compared to theoretical predictions and also to Monte
Carlo (MC) models .

\section{Theoretical Predictions}

The cross sections for isolated photon production  have been calculated to order
 ${O} {(\alpha^3\alpha_s)}$  by Gehrmann-De Ridder et
 al. (GGP)~\cite{GGP}.  In the approach of GGP, three contributions to the scattering 
cross-section  for $ep \rightarrow e\gamma X$ are considered at
leading order ($\alpha^3$) in the electromagnetic coupling. One of
these contributions comes from the radiation of a photon from the
quark line (called QQ photons) and a second from the radiation from
the lepton line (called LL photons). The third contribution is an 
interference term between photon emission from lepton and quark lines,  called LQ photons by
GGP.  In this analysis the contribution of the LQ term is reduced to a
negligible level by combining $e^+p$ and $e^-p$ data (the LQ term changes sign when $e^-$ is replaced by
$e^+$) and by requiring that the outgoing photon is well separated from both outgoing electron and
quark. Therefore this analysis took into account LL and QQ photons
only.

A calculation based on QED contributions to the parton distributions has been made by
Martin et al. (MRST)~\cite{MRST}.  
In the approach of MRST a resummed
version of the LL contribution is calculated. A photon parton component of the proton  arises as a direct consequence 
of including QED corrections in parton distribution functions. This
leads to a DGLAP-resummed enhancement of the LL prediction relative to that of GGP due to the inclusion of QED  Compton scattering, $\gamma_{p} e \rightarrow \gamma e$, where the $\gamma_{p}$ is a constituent of the proton. A measurement of the isolated high-energy photon  production cross section 
 therefore provides a direct constraint on the density of the
photon within the proton. The non-collinear QQ component is not included in the MRST
model 
and so the transverse momentum of
 the scattered electron is expected to balance approximately that of the isolated
    photon. The  analysis presented here does not impose this
    constraint.

\section{Data samples and Monte Carlo event simulation}
\label{sec:samples}
\label{sec-mc}
The measurements are based on a data sample corresponding
to an integrated luminosity of $320\pm8\,\mathrm{pb}^{-1}$, 
 taken between 2003 and 2007 with the ZEUS detector at HERA. The
 sample is a sum of $131\pm3  \,\mathrm{pb}^{-1}$ of $e^+p$ data and
 $189\pm 5 \,\mathrm{pb}^{-1}$ of $e^-p$ data with centre-of-mass energy $\sqrt{s}=318\,\mathrm{GeV}$\footnote{Hereafter 'electron' refers both
  to electrons and positrons unless specified.}.

DIS events with QQ photon emission were simulated by the MC program {\sc Pythia} 6.416
\cite{pythia}.  The LL photons radiated at large angles from 
the incoming or outgoing electron were simulated using the  generator
 {\sc Djangoh6} \cite{djangoh}, an interface to the MC program  {\sc Heracles} 4.6.6 \cite{heracles}; higher-order QCD effects were simulated  using the 
colour dipole model of {\sc Ariadne } 4.12\cite{ariadne}. Hadronisation  of 
the partonic final state was performed by {\sc Jetset}
\cite{jetset}. These samples were used to study the
event-reconstruction efficiency as well as providing the MC templates
for the signal extraction in Section \ref{sec:extr} and the MC predictions seen in
Section \ref{sec:results}.

The NC DIS background used in Section
\ref{sec:extr} was simulated using {\sc Djangoh6}, within the same 
framework as the LL events.

\section{Event selection and reconstruction}
\label{sec:selec}
The 
 kinematics were reconstructed from the scattered electron\footnote{The definitions of $Q^2$, $W_X$ and $x$ as reconstructed using the scattered electron are $Q^2=-(k-k')^2$, $W^2_X= (P + k -k' - p_{\gamma})^2 $ and $x=Q^2/(2P\cdot(k-k'))$ where $k$ ($k'$) is the four-momentum of the incoming (outgoing) lepton, $p_{\gamma}$ is the four vector of the outgoing photon  and $P$ is the four-momentum of the incoming proton.} and the kinematic region $10 <Q^2< 350\,\mathrm{GeV^2}$
 was chosen. 

 Scattered-electron candidates were and required to lie within
the polar angle range $139.8^{\circ}< \theta_e < 171.9^{\circ}$, to
ensure that they were well measured in the rear calorimeter (RCAL),  
and have energy, $E'_e$, greater than $10\,\mathrm{GeV}$. To reduce
backgrounds from non-$ep$ collisions and suppress deeply virtual compton scattering
\cite{dvcs} to a negligible level additional cuts were
made on the reconstructed vertex position, tracks and
energy sums.

 Photon candidates were identified
 as CAL energy flow objects (EFOs \cite{efos}) with reconstructed transverse energy within 
the range \mbox{$4 <E_T^{\gamma}<15\,\mathrm{GeV}$} and
pseudorapidity 
within the range $-0.7 < \eta^{\gamma}
  < 0.9$.  It was also required that at least $90\%$ of the reconstructed energy was measured in the
  barrel electromagnetic calorimeter (BEMC) and that the photon candidate was well isolated from reconstructed
  tracks.

Jet reconstruction was performed on all EFOs in the event, including the photon candidate, using the $k_T$ cluster algorithm \cite{kt} in the longitudinally 
invariant inclusive mode \cite{longinv}.  Further isolation was imposed by requiring that the 
energy of the photon-candidate EFO contributed at least $90 \%$ of the
 total energy of the jet to which it was assigned.


\section{Photon signal extraction}
\label{sec:extr}

The event sample selected according to the criteria in Section \ref{sec:selec} is dominated by background events, most notably neutral current (NC) DIS events where 
a genuine electron candidate is found in the RCAL and neutral mesons such as
 $\pi^0$ and $\eta$ decay to photons producing photon-candidate EFOs in the 
BEMC.  The photon signal
was extracted from the background using BEMC energy-cluster
shapes. Two shape variables were considered:

\begin{wrapfigure}{hr}{0.5\columnwidth}
\centerline{\includegraphics[width=0.45\columnwidth]{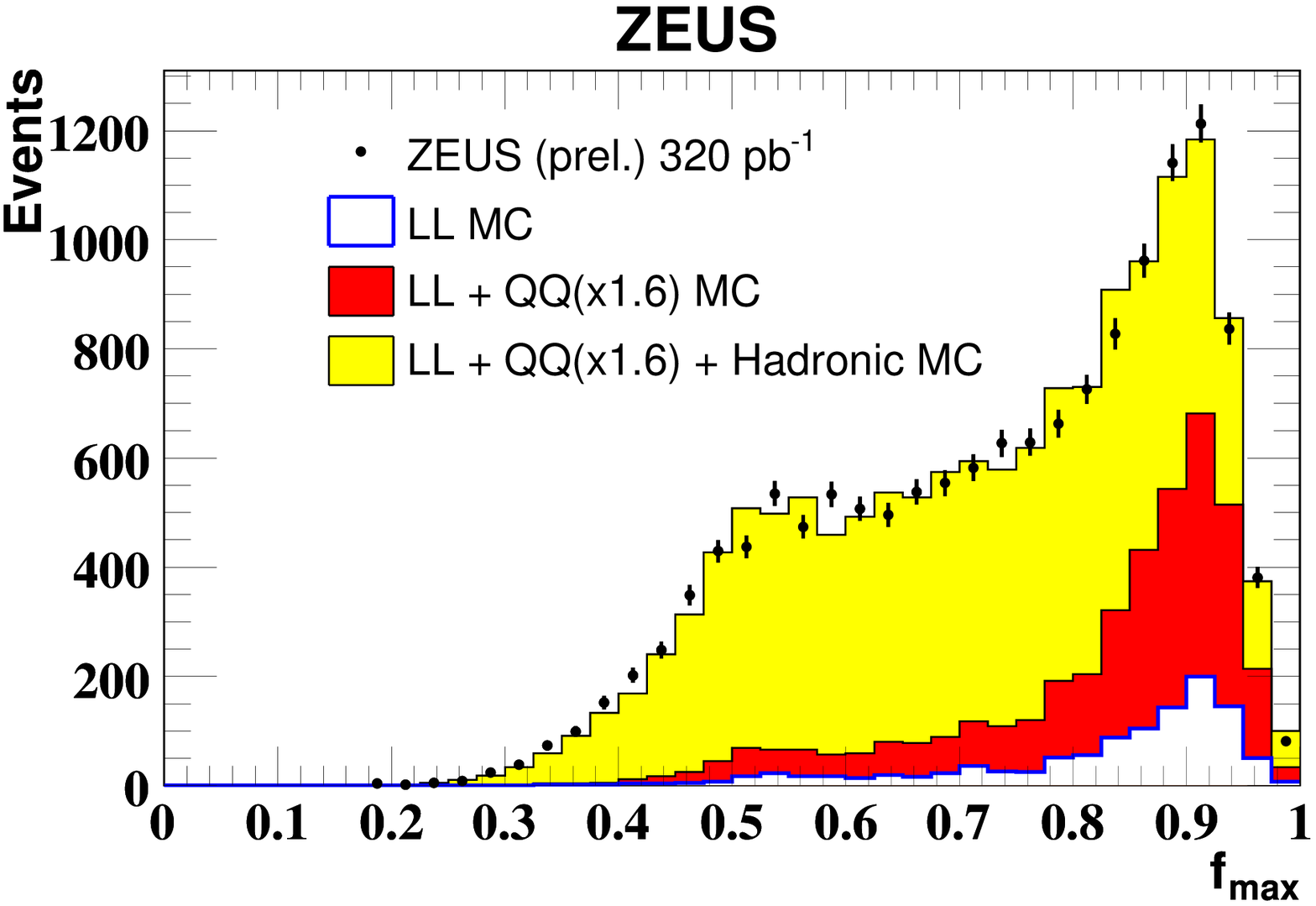}}
\centerline{\includegraphics[width=0.45\columnwidth]{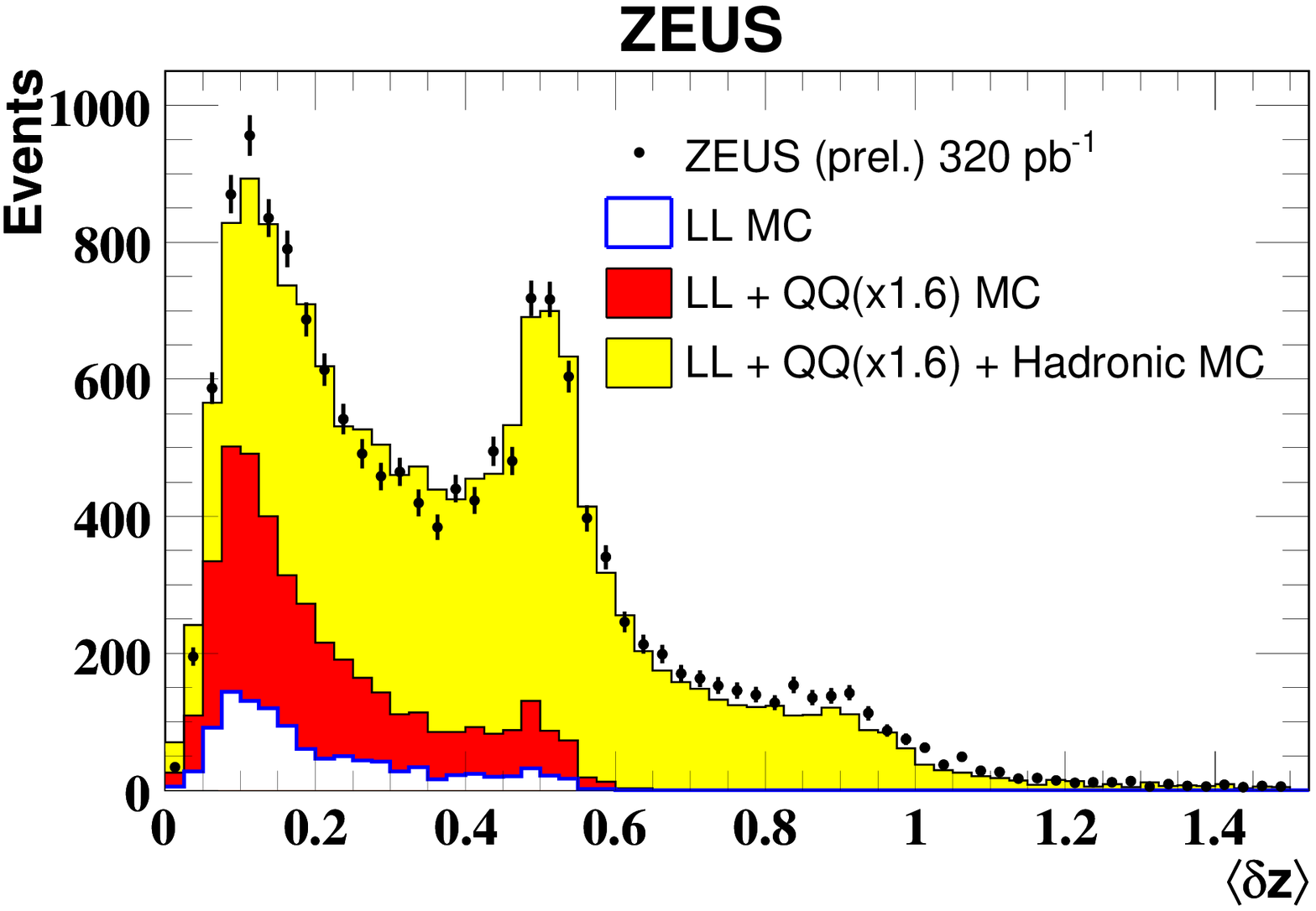}}
\caption{$f_{\mathrm{max}}$ and $\langle \delta Z\rangle$}
\label{fig:showers}
\end{wrapfigure}

\begin{itemize}
\item the variable  $\langle \delta Z\rangle= \frac{\sum \limits_i E_i|Z_i-Z_{\mathrm{cluster}}| }{W_{\mathrm{cell}}\sum \limits_i E_i}$, where  $Z_{i}$ is the $Z$ position of
 the centre of the $i$th cell,  $Z_{\mathrm{cluster}}$ is the centroid
 of the cluster, $W_{\mathrm{cell}}$ is the width of the cell in the $Z$ 
direction, $E_i$ is the energy recorded in the cell  and the sum runs over all BEMC 
cells in the EFO; 
\item the ratio of the highest-energy BEMC cell in the EFO to its total EMC energy, $f_{\mathrm{max}}$.
\end{itemize}


The distributions of $\langle \delta Z \rangle$ and $f_{\mathrm{max}}$ in data 
and MC are shown in Fig.~\ref{fig:showers}. 
 The $\langle \delta Z \rangle$ distribution exhibits a double-peaked structure with
the first peak at $\approx 0.1$, associated with the signal, and a
second peak at $\approx 0.5$, dominated by the $\pi^0 \rightarrow
\gamma \gamma$ background. The $f_{\mathrm{max}}$ distribution shows a single peak at
$\approx 0.9$ corresponding to the photon signal, and presents a ``shoulder'' extending down to $\approx 0.5$, which is dominated by the hadronic background.

The number of isolated-photon events contributing to Fig. \ref{fig:showers} and in each cross-section bin was determined by a
 $\chi^2$ fit to the $\langle \delta Z \rangle$ distribution in the range $0<\langle \delta Z \rangle < 0.8$ 
using MC-derived  templates for the LL and QQ signal distributions and
the background as described in Section \ref{sec:samples}. In
performing the fit, the LL contribution was kept constant at its  MC
predicted value and the other components were varied.  The final cross
sections were derived from the fitted events per bin using the MC
detailed in Section \ref{sec:samples} and the bin-by-bin unfolding method.






\section{Results and Conclusions}
\begin{figure}[h]
\includegraphics[width=0.5\columnwidth]{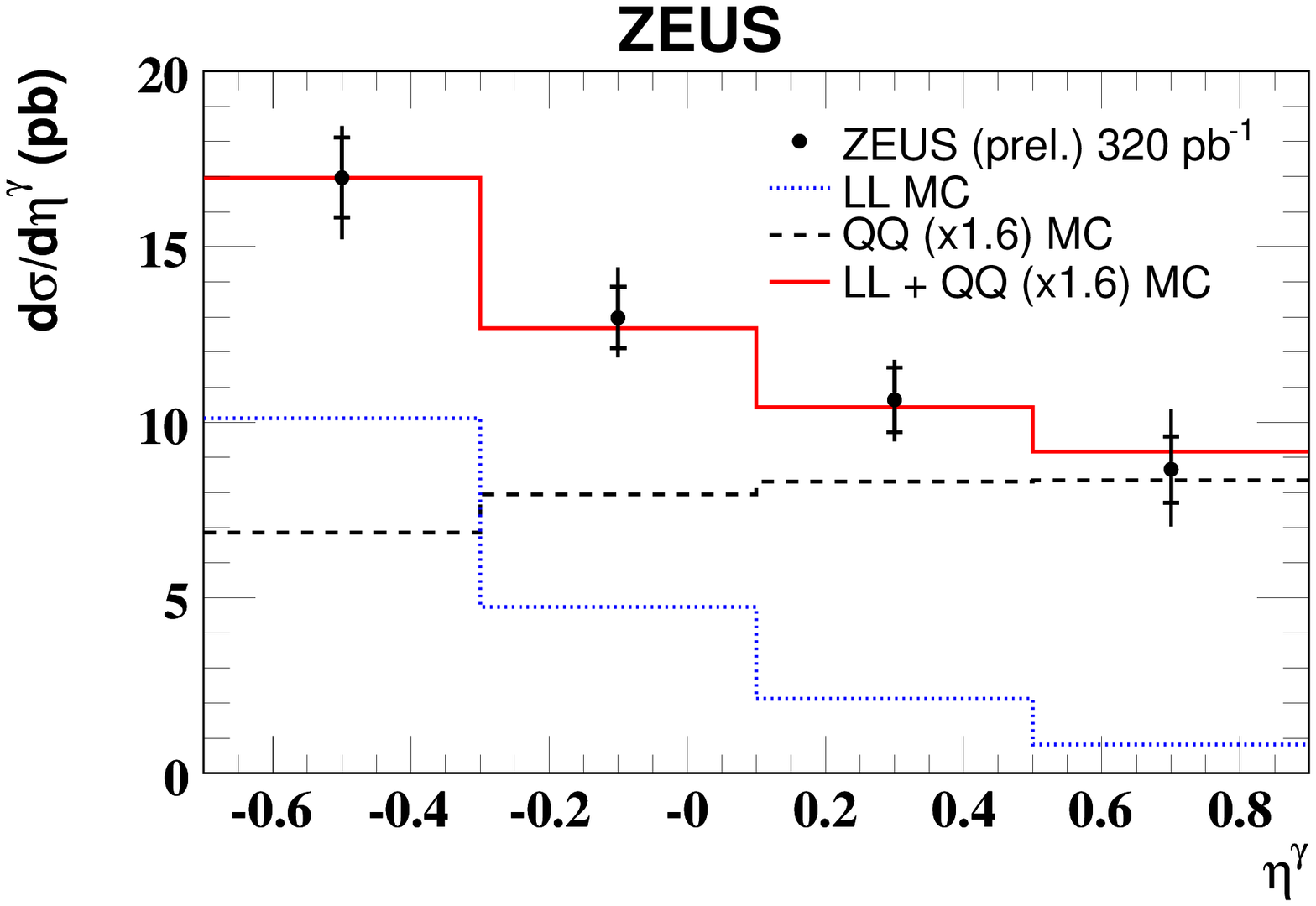}
\includegraphics[width=0.5\columnwidth]{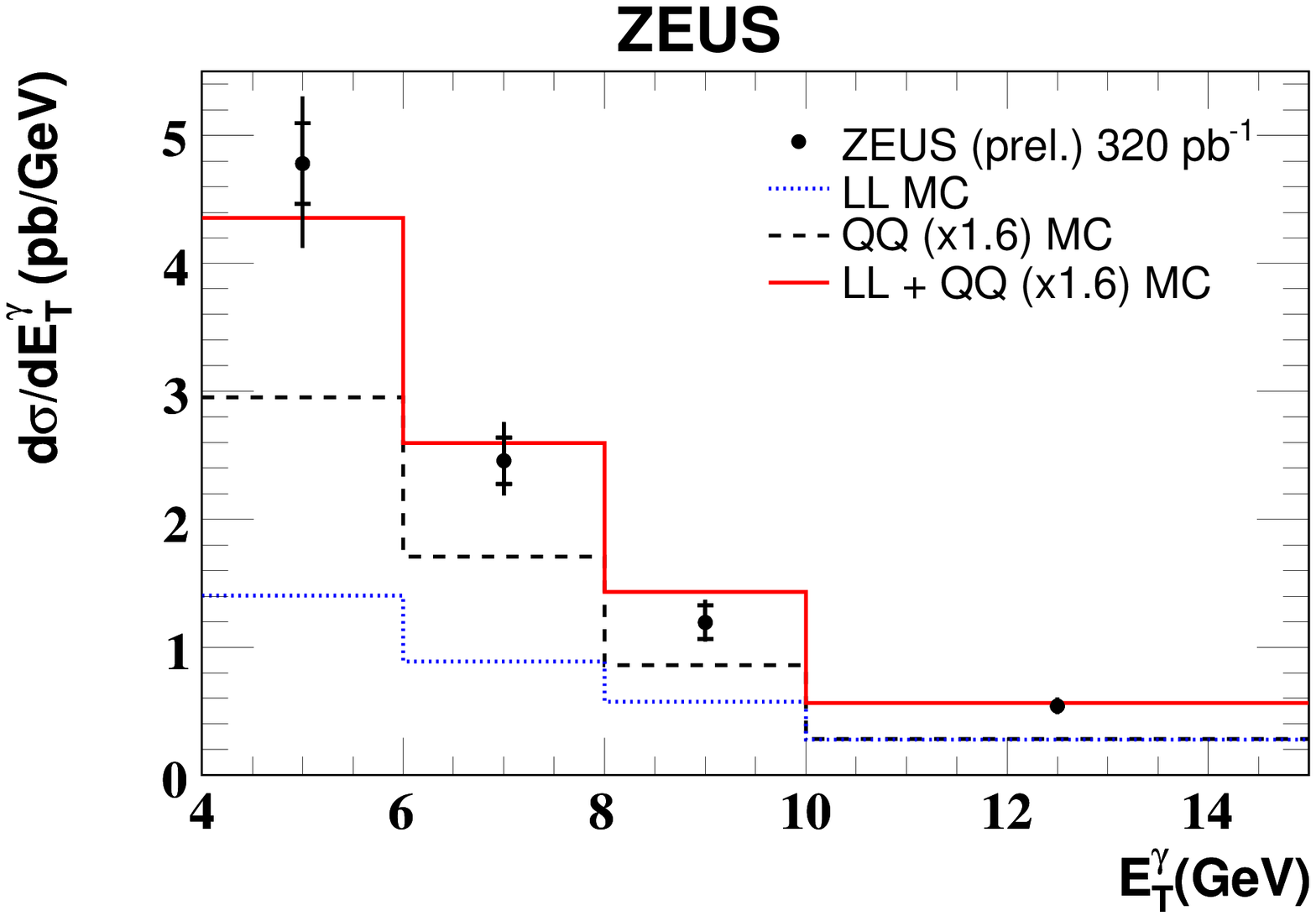}
\includegraphics[width=0.5\columnwidth]{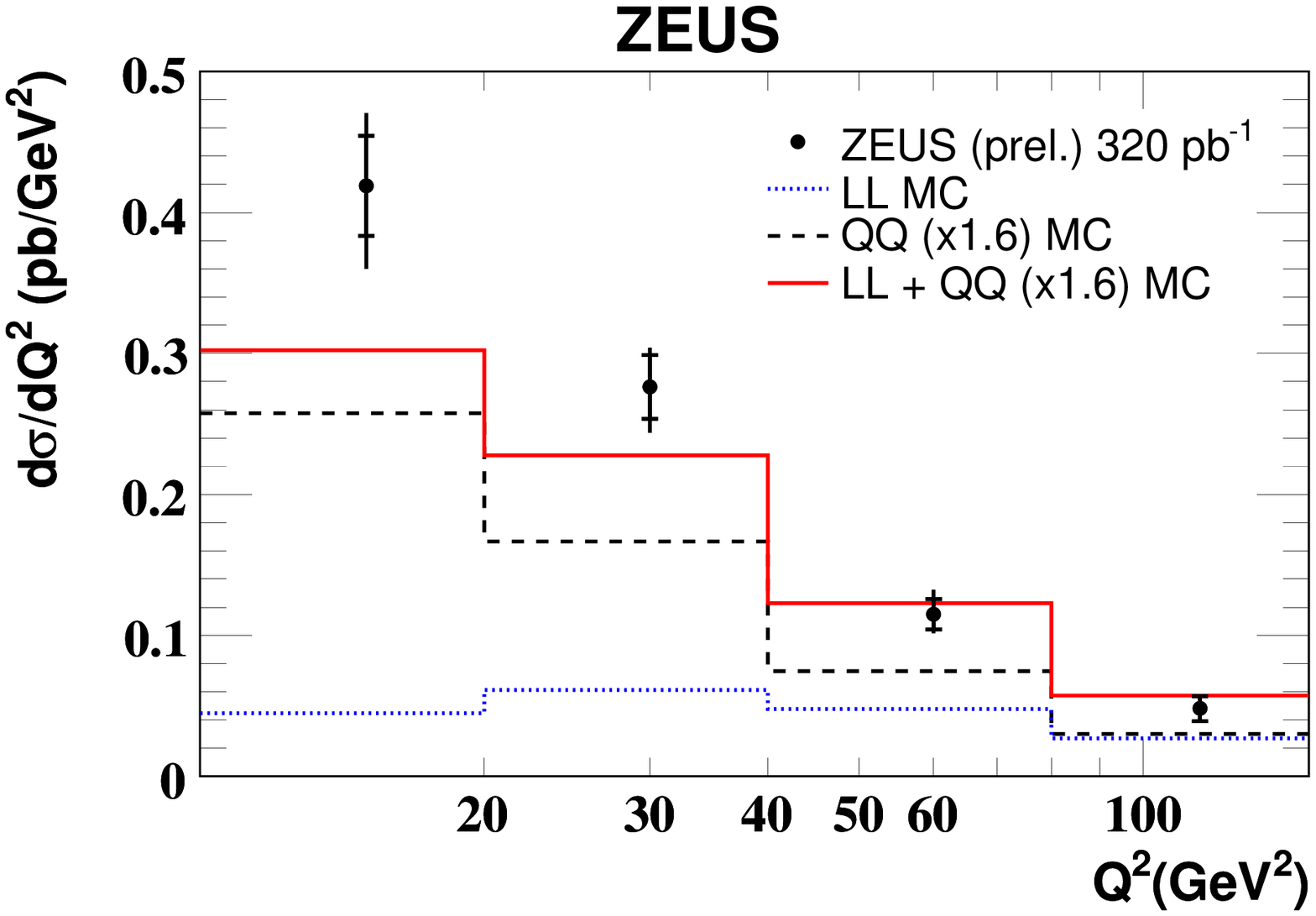}
\includegraphics[width=0.5\columnwidth]{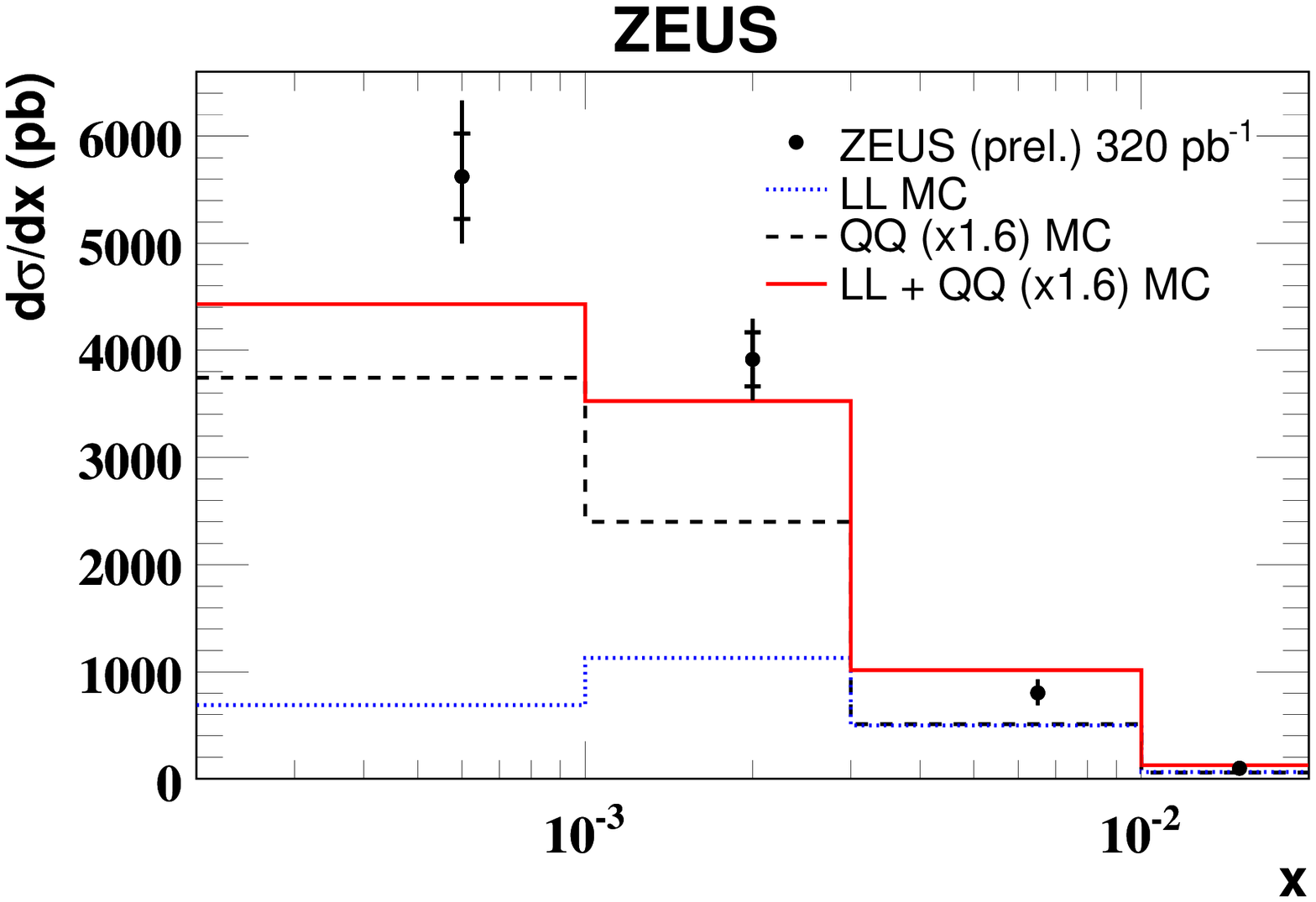}
\caption{\small 
Isolated photon differential cross sections compared to Monte Carlo
predictions.}
\label{fig:xsecmc}
\end{figure}

\begin{figure}[h]
\includegraphics[width=0.5\columnwidth]{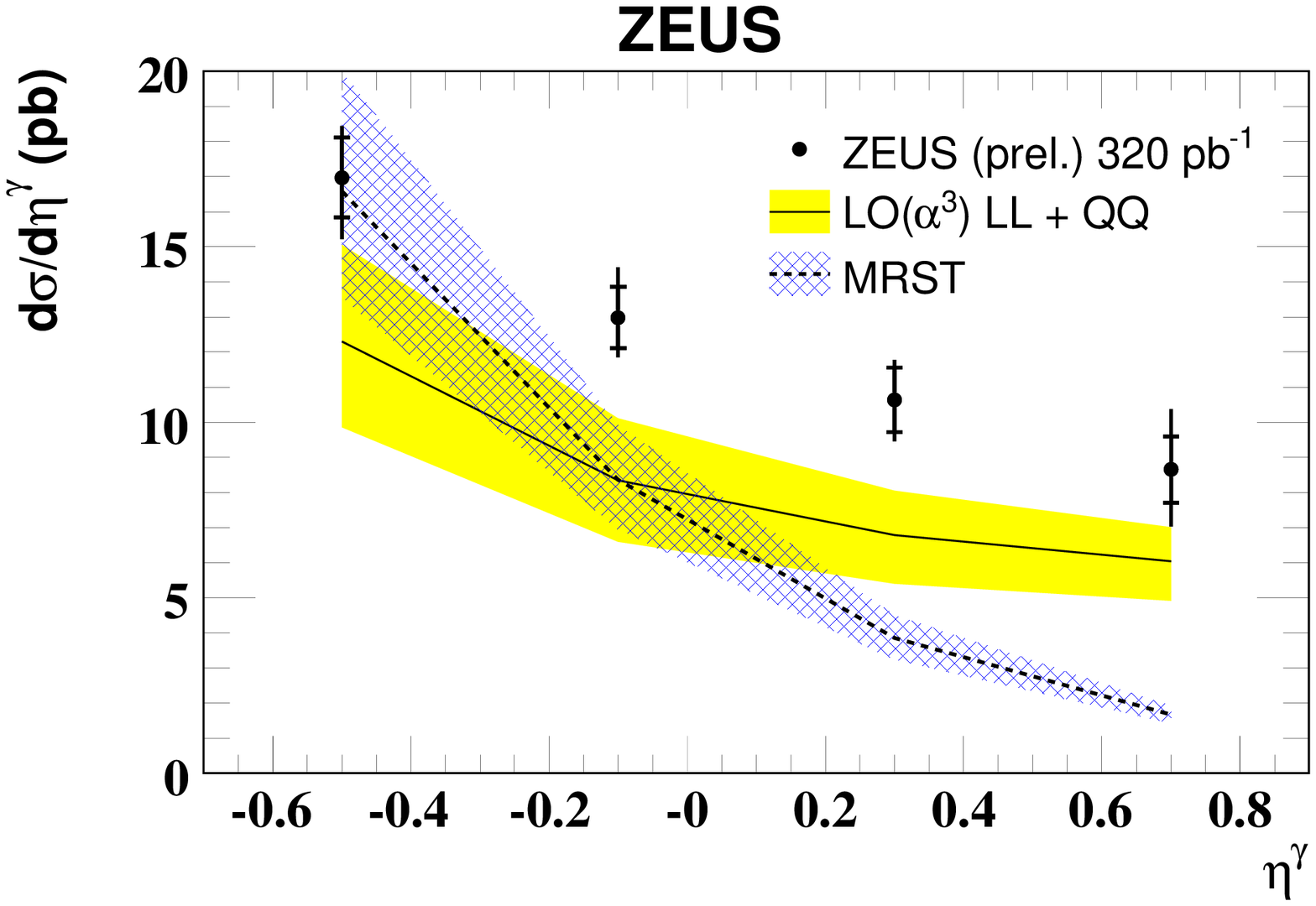}
\includegraphics[width=0.5\columnwidth]{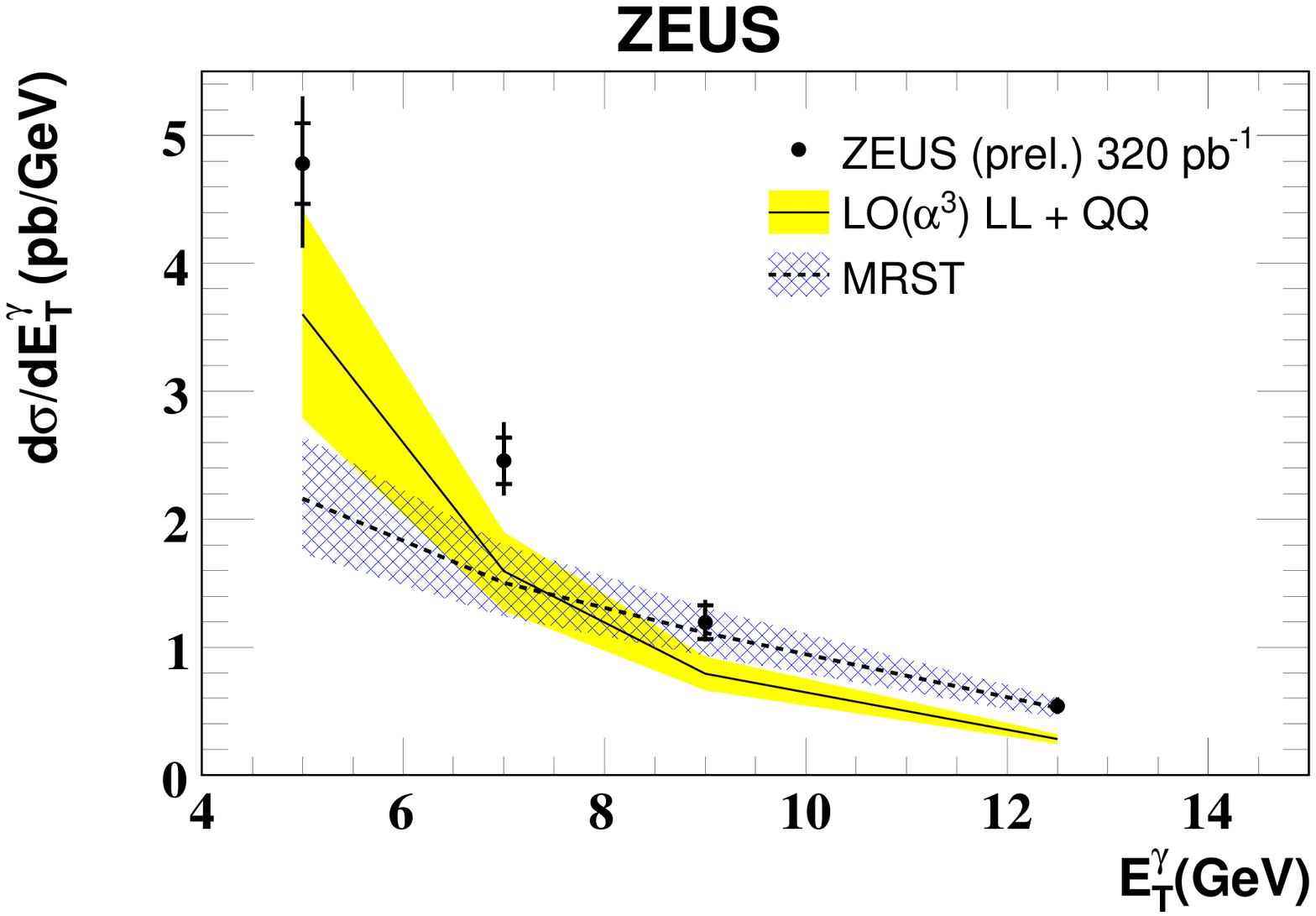}
\includegraphics[width=0.5\columnwidth]{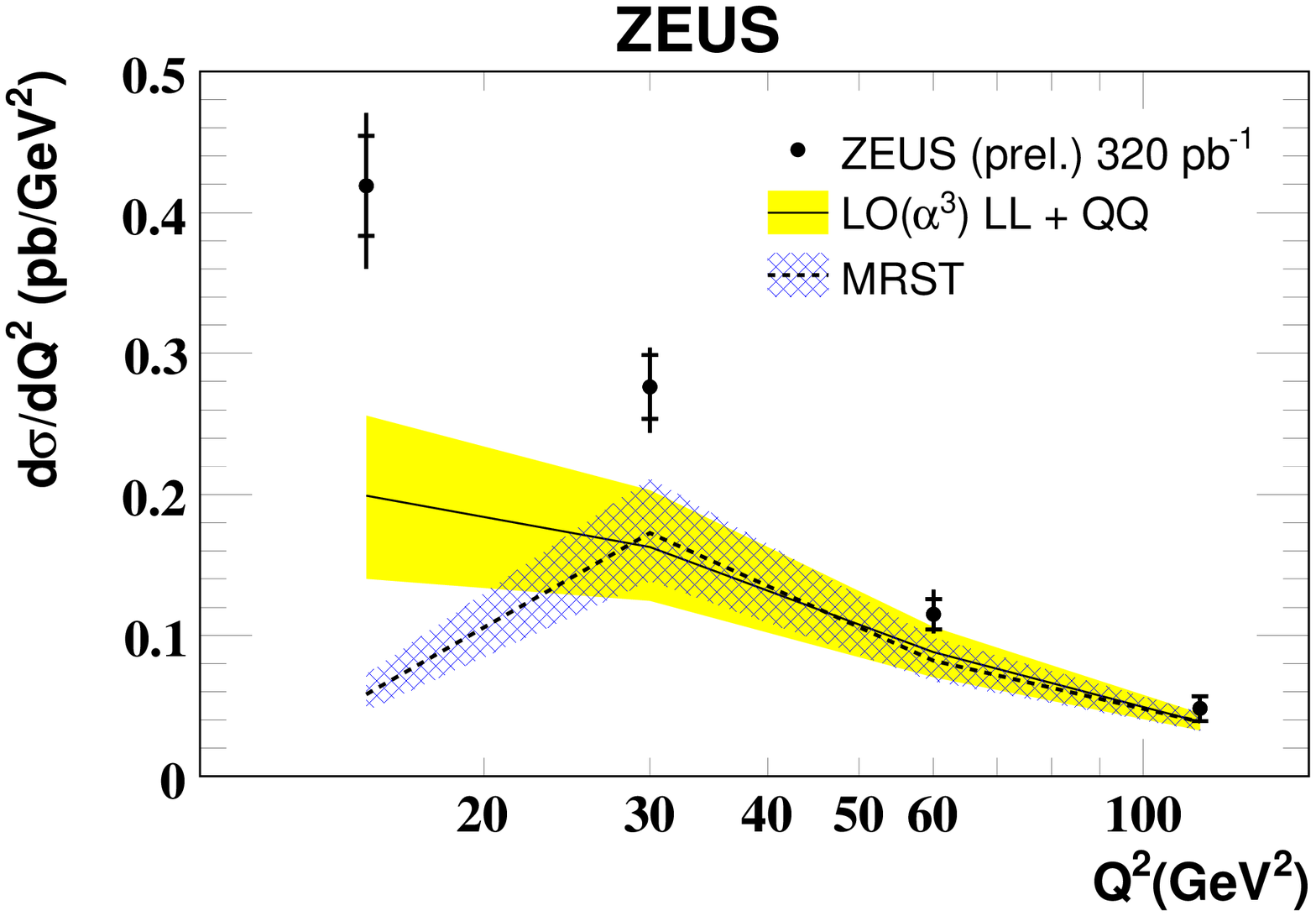}
\includegraphics[width=0.5\columnwidth]{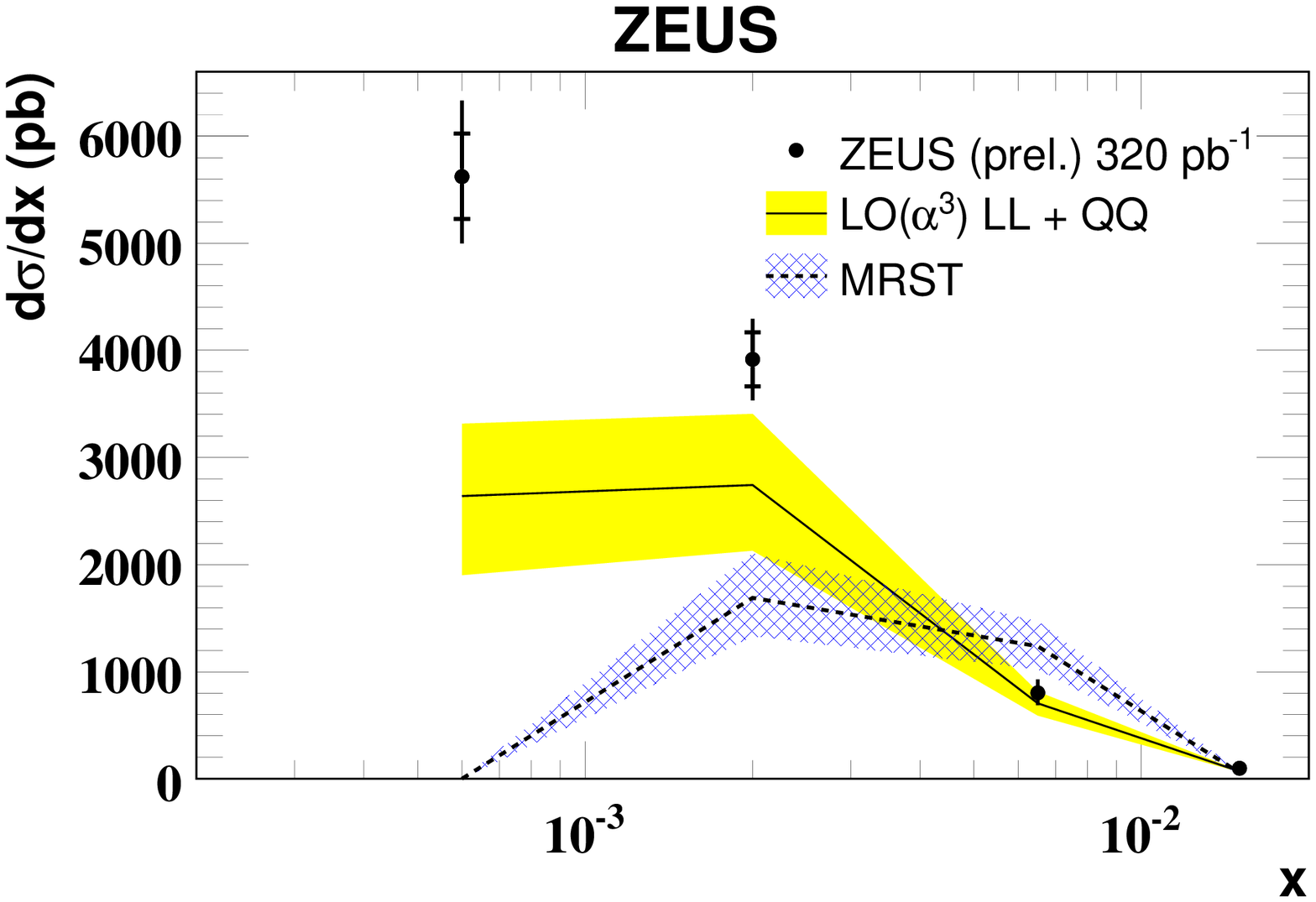}
\caption{\small 
Isolated photon differential cross sections compared to theoretical
calculations.}
\label{fig:xsectheory}
\end{figure}

\label{sec:results}


The differential cross sections as a function of  $E_T^{\gamma}$, $\eta^{\gamma}$, $Q^2$
and $x$ are shown  in Fig. \ref{fig:xsecmc}.  The predictions for the sum of the
expected LL contribution from {\sc Djangoh6} and a factor of
approximately 1.6 times the expected $QQ$ contribution from {\sc
  Pythia} agree well with the measurements, except for an 
excess at the lowest $Q^2$ (and associated lowest $x$). 

The differential cross sections as a function of  $\eta^{\gamma}$,
$E_T^{\gamma}$, $Q^2$ and $x$ are compared to theoretical predictions
in Fig. \ref{fig:xsectheory}.   The theoretical
 predictions from GGP describe the shape of the 
$E_T^{\gamma}$ and $\eta^{\gamma}$ distributions well, but the central value
 of the predictions typically lies $30 \%$ below the measured cross sections.
 The calculations fail to reproduce the shape in $Q^2$.
As with the MC comparison, the measured cross section is larger than
     the theoretical prediction; this is  reflected in an excess of data over theory at low-$x$.
 The MRST predictions fail to reproduce the shapes
of the measured differential cross sections. However they agree with the measurements at large values of $Q^2$ and $x$, for backward $\eta$ and for high values of $E_T$ where the the LL cross section is expected to be a larger fraction
       of the total.

\section{Acknowledgements}
We would like to thank T.~Gehrmann, W.J.~Stirling and R.~Thorne for their
predictions and many useful discussions.  We would also like to thank
DESY directorate for their support and the members of the ZEUS collaboration whose effort over many years has
made measurements such as this possible.

\begin{footnotesize}

\end{footnotesize}


\end{document}